# Device for Underwater Laboratory Simulation of Unconfined Blast Waves

Elijah Courtney[1], Amy Courtney[1], and Michael Courtney[1,a]

[1]BTG Research, 9574 Simon Lebleu Road, 70607, Lake Charles, Louisiana, USA

Shock tubes simulate blast waves to study their effects in air under laboratory conditions; however, few experimental models exist for simulating underwater blast waves that are needed for facilitating experiments in underwater blast transmission, determining injury thresholds in marine animals, validating numerical models, and exploring mitigation strategies for explosive well removals. This method incorporates an oxy-acetylene driven underwater blast simulator which creates peak blast pressures of about 1860 kPa. Shot-to-shot consistency was fair, with an average standard deviation near 150 kPa. Results suggest peak blast pressures from 460 kPa to 1860 kPa are available by adjusting the distance from the source.

Oil well demolitions using explosives have increased in recent years.[1] Blast waves can cause injuries apart from projectiles or impacts,[2] known as primary blast injuries. Blast waves from demolitions are injuring and even killing many different marine organisms.[3] Underwater blast effects have motivated experiments regarding blast injury thresholds of marine life,[4,5] underwater personal armor testing, blast transmission through water, and testing numerical models of blast waves in water in addition to experiments regarding underwater blast injury thresholds of humans and surrogate animals.[6,7] However, there are few available methods for simulating underwater blast waves in a laboratory environment.

Shock tubes are used to simulate air blast under laboratory conditions,[8,9] but there are few devices simulating blast waves in water.[10] The device used by Deshpande can create pressure waves with peak pressures ranging from 10 to 70 MPa with a duration of between 12 and 60 ms. Possible uses include small scale blast wave transmission testing of armor and other materials, but it is unclear how this device would be scaled up to test larger samples, applied to test blast injury thresholds in aquatic organisms, or validate computational models hoping to achieve biofidelic representations of larger animals.

Previous work showed that a modular, oxy-acetylene based shock tube produced realistic blast waves in air[8] with peak pressures up to about 5 MPa.[11] High blast pressures are desirable to more accurately simulate pressures created by underwater explosions.[3] The present study investigates a new approach to simulating underwater blast waves with an oxy-acetylene based device. This approach employs a Shchelkin spiral priming section which increases the turbulent flow of the deflagration wave, thus increasing its speed and pressure, facilitating a deflagration to detonation transition (DDT).

The priming section is coupled to an underwater tube oriented vertically in the water so the simulated blast wave moves outward from the center. Since blast waves decrease with distance, this geometry allows experimenters some control over blast exposure levels for testing injury thresholds in marine life and underwater blast transmission, and also for testing computational models, candidate armor materials, and other mitigation strategies.

A 30.5 cm long 2.54 cm diameter polyethylene tube with a wall thickness of about 0.07 mm was secured over the end of the priming section, and both were filled with a stoichiometric mixture of oxygen and acetylene following the method of Courtney et al.[11] Reaction of the oxy-acetylene mix was initiated by an impact to the priming compound. The priming section was a 60.7 cm long and 16 mm inner diameter machined steel tube with grooves of a depth of 0.36 mm, as in Courtney et al.[11]

Figure 1 illustrates the experimental arrangement. The plastic bag was placed underwater in a round container 64.0 cm high, 161.5 cm wide, 175.3 centimeters long. For experiments in blast wave transmission, samples may be placed between the blast source and pressure sensor. For experiments of blast injury in specimens, the specimen may be placed the proper distance from the source for the desired exposure level.

To characterize the device, two piezoelectric pressure sensors were placed. One sensor (PCB Piezotronics 102B06) was 22.9 cm or 30.5 cm from the center of the plastic tube containing the mixture of fuel and oxygen, and the other sensor (PCB Piezotronics 113B24) was 7.62 cm or 15.2 cm away from central axis of the simulated blast source. Pressure sensors were both placed at a depth of 15.2 cm under the surface of the water for each trial. (If a test specimen extends more than a few cm vertically, experimenters should ensure that the blast wave is relatively constant from top to bottom. Also, blast pressures should be checked if samples are tested at different depths.)

a) Author to whom correspondence should be addressed: Michael_Courtney@alum.mit.edu

The sensor faces were perpendicular to the blast wave direction of travel. Five trials for each distance were recorded with pressure sampled at 2 MHz via cables which connected the pressure transducer to a signal conditioning unit (PCB 842C) which produced a voltage output, which was digitized with a National Instruments USB-5132 analog to digital converter. Digitized voltage was converted to pressure using the calibration certificate provided by the manufacturer of the pressure sensor.

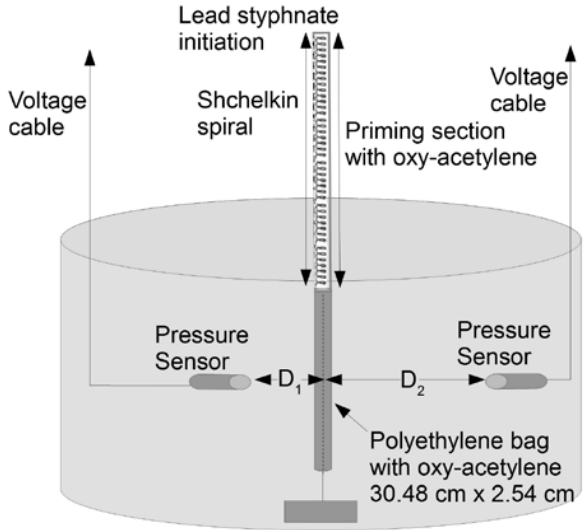

FIG 1: Diagram of underwater blast wave simulator.

Table I shows the average peak pressure at each distance and the standard error of the mean. Peak pressure decreased over distance, as expected. The duration of the blast event was about 0.05 ms for each distance; an evident reflection was observed for most trials. Using a larger tank or a tank whose boundaries transmitted or absorbed the blast wave may be required to meet some experimental needs. Alternatively, tank geometry may be used to direct undesirable reflected waves away from the object under test. It may also be useful in some experiments to expose the object under test to both the direct and reflected blast waves.

TABLE I: The mean peak pressure of five trials and standard error of the mean in kPa for each distance measured.

| Distance | Mean Peak Pressure (kPa) | Standard Error of the Mean (kPa) |
|---|---|---|
| 7.6 cm | 1856 | 80 |
| 15.2 cm | 1045 | 97 |
| 22.9 cm | 896 | 62 |
| 30.5 cm | 458 | 25 |

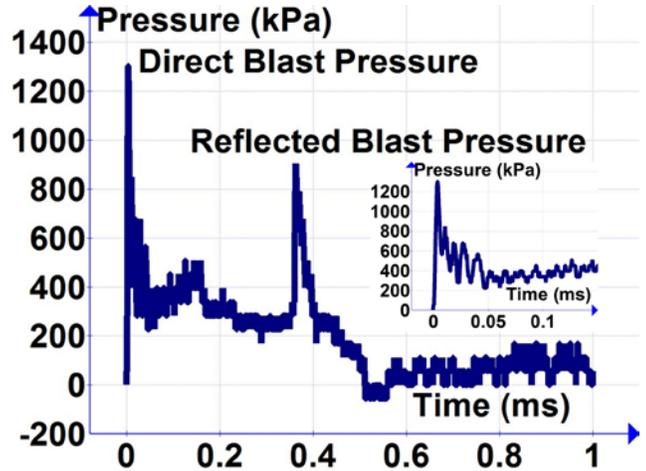

FIG 2: A typical graph of the blast wave measured at 15.2 cm. Note the secondary peak created by the reflected blast pressure at 0.36 ms.

Figure 2 shows a pressure-time curve measured at 15.2 cm. The shape of the curve is typical. Figure 2 also shows a clear reflection at about 0.36 ms, with an amplitude of about 900 kPa. Unlike free-field air blast where pressure-time curves are well modeled by a Friedlander waveform, underwater blasts are not well modeled by a simple mathematical function, even though they are often likened to a decaying exponential.[10, 12] The early time profile (Figure 2, inset) shows some smaller component peaks which may be reflections from the top and bottom of the water. These shapes are not unlike the underwater blast wave simulator of Deshpande[10] and underwater blast measurements from offshore well removals.[13] Thus, while blast waves from this device may not be as ideal as well-formed air blast waves from shock tubes, they are reasonable for their intended purpose.

The simulated blast wave that originates in the cylindrical oxygen-fuel mixture travels outward in all directions and first reaches the sensor by the direct path through the water, thus traveling 15.2 cm in the trial shown in Figure 2. The component of the wave that propagates directly away from the sensor strikes the side of the tank located 87.7 cm from the point of origin (center of the tank) and is reflected and partially focused back toward the center of the tank. This reflected wave then reaches the sensor after traveling 15.2 cm more for a total travel distance of 190.5 cm. Thus, the reflected wave traveled 175.3 cm more than the direct wave in an additional time of 0.36 ms, suggesting a wave propagation speed of 4869 m/s, which is consistent with an underwater blast wave.

Figure 3 shows the mean peak pressure for each distance. The mean peak pressure is fit to a power law, because physical intuition suggests that point sources fall off as $1/r^2$ and line sources fall off as $1/r$. Fitting to the power law, $P(r) = 1000(r/r_1)^b$, yields $r_1 = 14.84 \pm 0.72$ cm and $b = -0.955 \pm 0.095$. The adjustable best fit parameter $r_1$ corresponds to the distance where $P = 1$ MPa.

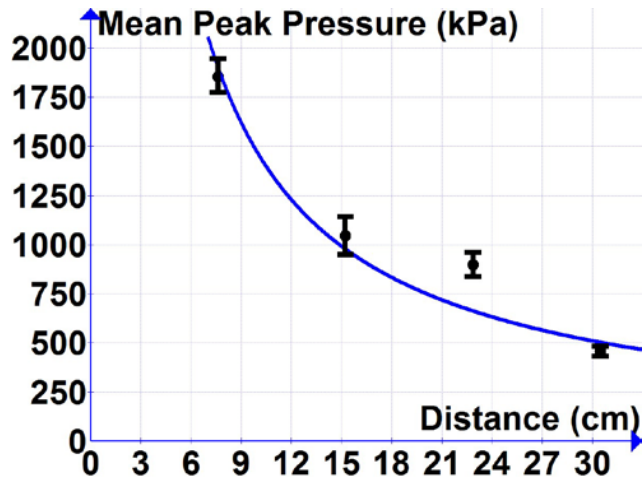

FIG 3: The average peak pressure vs distance fit to a power law. Error bars indicate the standard error of the mean.

Blast waves both traveled rapidly and fell off quickly in water. The closest distance measured had a mean peak pressure of about 1850 kPa, while the farthest distance had a mean peak pressure of about 460 kPa, with a mean wave speed of nearly 4870 m/s. The results from each set of experiments were fairly repeatable, with standard deviations ranging from 9.7-21%. The blast wave has a steep shock front with a rapid initial decay.

It is of interest to know how blast waves act underwater, how water acts as a blast mitigator and the rate at which water attenuates a blast wave. Experimental devices can be used to test naval armor and to measure the blast tolerances of marine life. Future work may include testing of the air-water boundary as a blast mitigator. This would include a blast source located above the surface of the water, with the sensor directly below, to assess how the difference in acoustic impedance attenuates the blast wave. Alternatively, the blast source might be underwater, with a "curtain" of air bubbles between the source and the sensor.

One possible future study might employ a container of elliptical cross section. The blast source could be located at one focus of the ellipse and the object being tested would be at the other focus, thus magnifying the blast wave exposure. The object could thus be subjected to a variety of pressures, depending on how far from the focus the object was. Also, the object would be subjected to blast waves from all directions.

This experimental design required almost no specialized or expensive parts, and experiments are possible without the cost, liability, regulations, or bureaucratic approvals associated with laboratory use of high explosives. A possible limitation is an uneven distribution of oxygen-acetylene resulting from pressure changes with depth. This limitation can be mitigated by care to sufficiently inflate and seal the bag so that hydrostatic pressure does not cause the fuel-oxygen mix to escape from the bag or produce a noticeable decrease in bag diameter with depth. With due experimental care, the dominant effect will be limited to about 3% for the depths used here and 10% per meter for greater depths.

It is foreseeable that this technique could be used to generate more powerful blast waves or waves more uniform over larger spatial extents by using larger diameter and/or longer bags. This approach may prove advantageous over the alternative of creating simulated blast waves by high speed entry of projectiles.[14] Water entry by projectiles may be able to produce pressure transients more closely approximating a decreasing exponential; however, confined oxy-acetylene or other oxy-fuel mixtures may succeed in producing waves more uniform over larger test samples.

1. Marine Board of the National Research Council. "A focus on offshore safety: recent reports by the Marine Board of the National Research Council," Marine Board, National Research Council (1997).
2. I. Cernak, Z. Wang, J. Jiang, X. Bian, and J. Savic. "Ultrastructural and functional characteristics of blast injury-induced neurotrauma," The Journal of Trauma, Injury, Infection, and Critical Care 50(4), p.695-706 (2001).
3. D. Bagocius. "Underwater noise generated by the detonation of historical ordnance in the Baltic Sea, Lithuania: potential ecological impacts on marine life," BALTICA 26 (2), p.187–192 (2013).
4. J. T. Yelverton, D. R. Richmond, W. Hicks, K. Saunders, and E.R. Fletcher. "The Relationship between fish size and their response to underwater blast," Lovelace Foundation for Medical Education and Research (1975).
5. J.R. Nedwell, S.J. Parvin, and E. Harland. "Lethal and physical injury of marine mammals, and requirements for passive acoustic monitoring," Subacoustech Ltd. (2007).
6. E.R. Fletcher, J.T. Yelverton, and D.R. Richmond. "The thoraco-abdominal system's response to underwater blast," Lovelace Foundation for Medical Education and Research (1976).
7. D.R. Richmond, J.T. Yelverton, and E.R. Fletcher. "Far-field underwater-blast injuries produced by small charges," Lovelace Foundation for Medical Education and Research (1973).
8. A. Courtney, L. Andrusiv, M. Courtney. "Oxy-acetylene driven laboratory scale shock tubes for studying blast wave effects," Review of Scientific Instruments 83, 045111 (2012).
9. W.A. Martin. "A review of shock tubes and shock tunnels," General Dynamics, San Diego, CA, Convair Div, No. GDC-ZR-658-050, (1958).
10. V.S Deshpande, A Heaver, and N.A Fleck. "An underwater shock simulator," Proceedings of the Royal Society for Mathematical, Physical and Engineering Science (2006).
11. E.D.S. Courtney, A.C. Courtney, and M.W. Courtney. "Shock tube design for high intensity blast waves for laboratory testing of armor and combat materiel," Defence Technology (2014).
12. R.H Cole. "Underwater Explosions," Princeton University Press (1948).
13. J.G. Connor, Jr. "Underwater blast effects from explosive severance of offshore platform legs and well conductors," (No. NAVSWC-TR-90-532), Naval Surface Warfare Center, Silver Spring, MD (1990).
14. M. Lee, R.G. Longoria, and D.E. Wilson. "Ballistic waves in high-speed water entry," Journal of fluids and Structures, 11(7), p. 819-844 (1997).